\documentclass[showkeys,superscriptaddress,amsmath,amssymb,aps,prx,twocolumn,floatfix]{revtex4-1}

\usepackage[utf8x]{inputenc}
\usepackage{graphicx}
\usepackage{dcolumn}
\usepackage{bm}
\usepackage{xcolor}
\usepackage{amsmath}
\usepackage{chngcntr}
\usepackage{nicefrac}
\usepackage[colorlinks,citecolor = blue,linkcolor=blue,hyperindex,CJKbookmarks]{hyperref}

\begin{document}

\title{Using magnetic dynamics to measure the spin gap in a candidate Kitaev material}

\author{Xinyi Jiang}\thanks{These authors contributed equally to this work.}
\author{Qingzheng Qiu}\thanks{These authors contributed equally to this work.}
\affiliation{International Center for Quantum Materials, School of Physics, Peking University, Beijing 100871, China}
\author{Cheng Peng}\thanks{These authors contributed equally to this work.}
\affiliation{Stanford Institute for Materials and Energy Science, Stanford University and SLAC National Accelerator Laboratory, Menlo Park, California 94025}

\author{Hoyoung Jang}
\affiliation{PAL-XFEL, Pohang Accelerator Laboratory, POSTECH, Pohang, Gyeongbuk, 37673 Republic of Korea}
\affiliation{Photon Science Center, POSTECH, Pohang, Gyeongbuk, 37673 Republic of Korea}

\author{Wenjie Chen}
\author{Xianghong Jin}
\author{Li Yue}
\affiliation{International Center for Quantum Materials, School of Physics, Peking University, Beijing 100871, China}

\author{Byungjune Lee}
\affiliation{Max Planck POSTECH/Korea Research Initiative, Center for Complex Phase Materials, Pohang, Gyeongbuk, 37673 Republic of Korea}

\author{Sang-Youn Park}
\author{Minseok Kim}
\author{Hyeong-Do Kim}
\affiliation{PAL-XFEL, Pohang Accelerator Laboratory, POSTECH, Pohang, Gyeongbuk, 37673 Republic of Korea}

\author{Xinqiang Cai}
\author{Qizhi Li}
\affiliation{International Center for Quantum Materials, School of Physics, Peking University, Beijing 100871, China}

\author{Tao Dong}
\affiliation{International Center for Quantum Materials, School of Physics, Peking University, Beijing 100871, China}

\author{Nanlin Wang}
\affiliation{International Center for Quantum Materials, School of Physics, Peking University, Beijing 100871, China}
\affiliation{Collaborative Innovation Center of Quantum Matter, Beijing 100871, China}

\author{Joshua J. Turner}
\affiliation{Stanford Institute for Materials and Energy Science, Stanford University and SLAC National Accelerator Laboratory, Menlo Park, California 94025}
\affiliation{Linac Coherent Light Source, SLAC National Accelerator Laboratory, Menlo Park, CA 94720}

\author{Yuan Li}
\affiliation{International Center for Quantum Materials, School of Physics, Peking University, Beijing 100871, China}
\affiliation{Collaborative Innovation Center of Quantum Matter, Beijing 100871, China}

\author{Yao Wang}
\email{yao.wang@emory.edu}
\affiliation{Department of Chemistry, Emory University, Atlanta, GA 30322, USA}

\author{Yingying Peng}
\email{yingying.peng@pku.edu.cn}
\affiliation{International Center for Quantum Materials, School of Physics, Peking University, Beijing 100871, China}
\affiliation{Collaborative Innovation Center of Quantum Matter, Beijing 100871, China}

\date{\today}

\begin{abstract}

Materials potentially hosting Kitaev spin-liquid states are considered crucial for realizing topological quantum computing. However, the intricate nature of spin interactions within these materials complicates the precise measurement of low-energy spin excitations indicative of fractionalized excitations. Using Na$_{2}$Co$_2$TeO$_{6}$ as an example, we study these low-energy spin excitations using the time-resolved resonant elastic x-ray scattering (tr-REXS). Our observations unveil remarkably slow spin dynamics at the magnetic peak, whose recovery timescale is several nanoseconds. This timescale aligns with the extrapolated spin gap of $\sim$ 1\,$\mu$eV, obtained by density matrix renormalization group (DMRG) simulations in the thermodynamic limit. The consistency demonstrates the efficacy of tr-REXS in discerning low-energy spin gaps inaccessible to conventional spectroscopic techniques.

\end{abstract}

\maketitle

\section{Introduction}

The exactly solvable Kitaev model with honeycomb magnetic lattice geometry\,\cite{kitaev2006anyons} has garnered substantial interest in quantum spin liquid (QSL) research\,\cite{zhou2017quantum}. In this model, bond-dependent anisotropic spin interactions between adjacent spins lead to magnetic frustration and significant quantum fluctuations\,\cite{liu2020kitaev}, preventing the formation of long-range spin order even at zero temperature. As a consequence, its highly competing ground states form the Kitaev QSL\,\cite{ashkin1943propagation, balents2010spin}. This result offers a promising foundation for topological quantum computing due to the emergence of fractional Majorana excitations\,\cite{takagi2019concept}. Nonetheless, the Kitaev model represents an idealized theory. In reality, candidate Kitaev materials often exhibit complicated magnetic interactions, such as the isotropic Heisenberg interaction and off-diagonal spin exchange, aside from the Kitaev interactions\,\cite{liu2018pseudospin}. The combination of these interactions yields a diverse phase diagram including QSL and various ordered phases.

\begin{figure}[!t]
\centering
\includegraphics[width=\columnwidth]{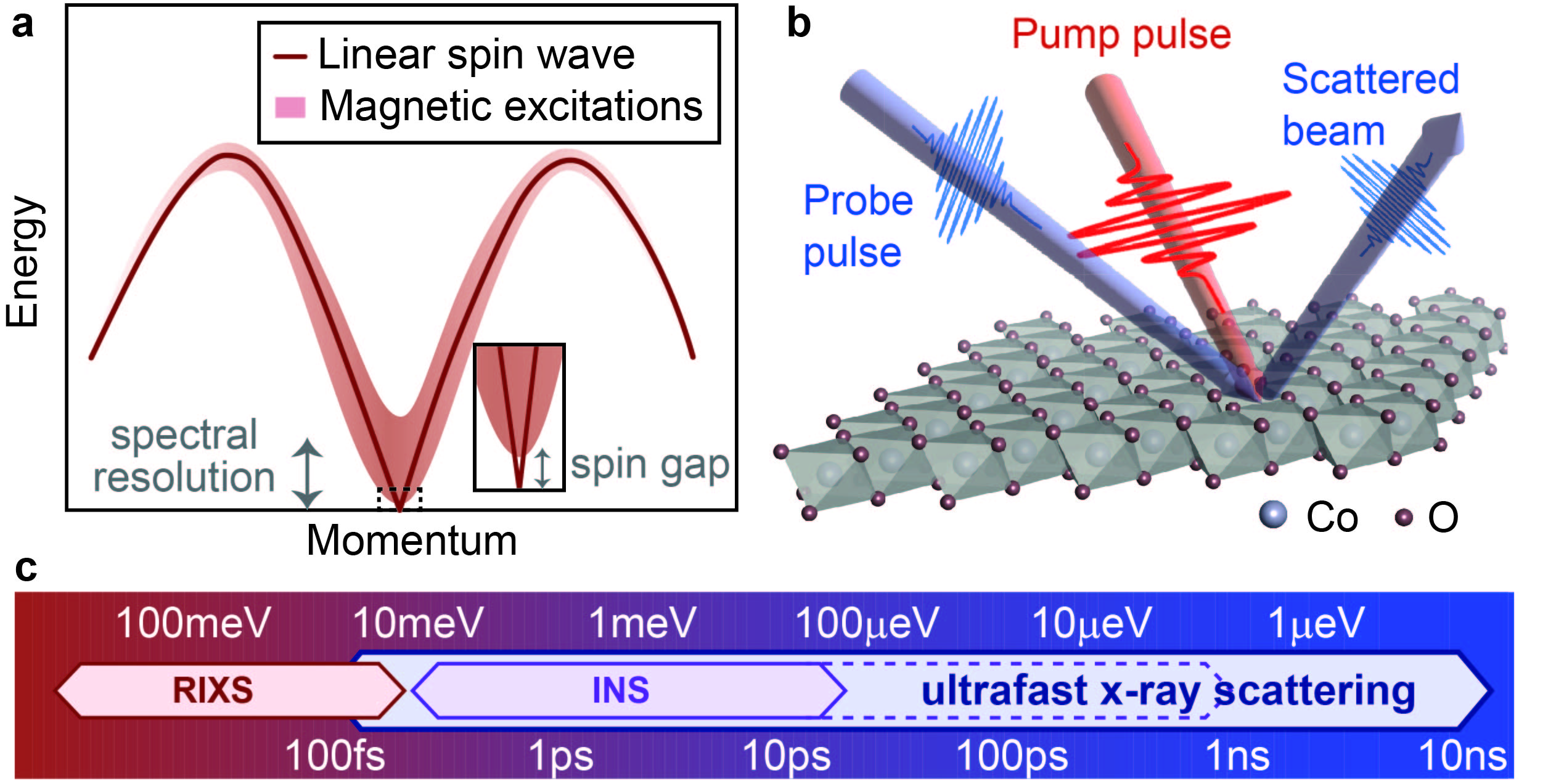}\vspace{-2mm}
\caption{{\bf Ultrafast characterization of low-energy spin gaps.} \textbf{a} Illustration of gapless spin-wave excitations (solid line) and the continuum in frustrated magnets.  A closer look at the low-energy region is provided in the inset. The arrows sketch the spectral resolution and the small spin gap. \textbf{b} Experimental setup for the trREXS, showing the optical pump (red) and x-ray probe (blue). \textbf{c} Comparison of energy resolutions of scattering spectra and their corresponding timescales. The red arrow sketches the range of energy resolutions accessible through RIXS; the purple arrow indicates the energy resolution commonly achieved by INS, with the dashed line representing resolutions that require additional effort. Ultrafast trREXS measurement can cover a broad energy resolution range based on the observed timescales.\vspace{-3mm}
\label{fig:cartoon}}
\end{figure}

Several transition-metal materials, including $\alpha$-RuCl$_3$\,\cite{PhysRevLett.119.227208} and H$_3$LiIr$_2$O$_6$\,\cite{H3LiIr2O6}, have been proposed as candidates for realizing the spin-$1/2$ Kitaev model through spin-orbital coupling (SOC) at the transition-metal centers. Apart from these, Na$_{2}$Co$_2$TeO$_{6}$ is distinguished as a promising candidate due to its more compact $3d$ orbitals which exhibit stronger spin couplings compared to the weakly localized $5d$ and $4d$ metals\,\cite{liu2020kitaev}. The SOC within $3d$ orbitals of each Co atom gives rise to an effective spin-$1/2$ configuration on the honeycomb lattice. At low temperatures, the system transitions from a paramagnetic state to a 2D antiferromagnetic ordered state at 31\,K, followed by the emergence of a three-dimensional (3D) ordered state below 26.7\,K\,\cite{PhysRevB.103.L180404,PhysRevLett.129.147202,PhysRevB.103.214447}. With frustration, the pronounced spin correlations result in competing magnetic ground states at low-temperature \,\cite{lefranccois2016magnetic, bera2017zigzag, PhysRevB.103.L180404,PhysRevLett.129.147202,PhysRevB.103.214447}. The intricate low-energy spin excitations, coupled with the challenges of limited spectral resolution, have impeded the precise determination of its spin gap.

More generally, characterizing small spin gaps is significant for identifying phases in quantum materials. Historical examples include the debates over whether various spin systems exhibit a gapped or gapless spin liquid, such as the initial variational calculations proposing a gapless Dirac QSL state in the spin-$1/2$ Heisenberg model on kagome lattices\,\cite{PhysRevLett.98.117205}, while DMRG studies suggested a gapped QSL state with distinct properties\,\cite{doi:10.1126/science.1201080}. Beyond spin liquids, the presence of spin gap signals non-trivial topological properties and thermal transport properties in low-dimensional materials\,\cite{spingapinTIs, PhysRevX.8.041028,PhysRevB.108.144402,PhysRevResearch.5.043110, PhysRevX.12.041031}. The ability to directly detect the spin gap is crucial to resolving these questions.
Thus, various conventional experimental techniques have been employed to characterize the spin gap, as shown in Fig.~\ref{fig:cartoon}\textbf{a}, including specific heat\,\cite{PhysRevLett.56.185}, 
thermal conductivity\,\cite{thermalconductivity,hong2023phonon}, electron spin resonance\,\cite{ESRNCTO,ESRspingap2}, nuclear magnetic resonance (NMR)\,\cite{NMRspingap}, inelastic neutron scattering (INS)\,\cite{PhysRevB.98.220402}, and resonant inelastic x-ray scattering (RIXS)\,\cite{PhysRevLett.109.157402}. However, detecting spin gaps smaller than microelectron volts by these traditional techniques remains a significant challenge, limited by the lower bounds of measured energy and temperature, as well as energy resolution.

An alternative method for probing small-energy excitations has been proposed in the time domain. As shown in Fig.~\ref{fig:cartoon}\textbf{c}, the characteristic time is inverse to a dominant energy scale. Thus, a small spin gap, which is beyond the resolution of RIXS or INS measurements, reflects a relatively long timescale that can be discerned using pump-probe techniques. For example, time-resolved optical spectroscopy has successfully disentangled low-energy bosonic excitations through their distinct timescales\,\cite{10.1126/science.1216765}; time-resolved x-ray scattering spectroscopy has revealed a collective charge fluctuation with characteristic energy in the sub-meV range\,\cite{LBCOMM}; x-ray photon correlation spectroscopy has been utilized to reveal sub-meV antiferromagnetic domain fluctuations \cite{Nature4476871}.
In this scenario, ultrafast x-ray scattering emerges as a promising avenue to achieve microelectron volt energy resolution for a specific magnetic excitation by analyzing the corresponding finite-momentum dynamics in the time domain. 
By monitoring dynamics exceeding longer than several nanoseconds, one can access an energy resolution of sub-$\mu$eV scales, which is crucial for probing small spin gaps.

To this end, we employ time-resolved resonant elastic x-ray scattering (Tr-REXS) to reveal long-term magnetic dynamics at picoseconds to nanoseconds timescales in the Kitaev candidate material Na$_{2}$Co$_2$TeO$_{6}$, as sketched in Fig.~\ref{fig:cartoon}\textbf{b}. With the high-momentum resolution and time-resolved capabilities, we are able to directly investigate the fluence- and temperature-dependence of magnetic dynamics after pump. By a DMRG simulation of its model Hamiltonian, we further show that observed slow recovery dynamics reflects the small spin gaps in these types of magnetic materials. Thus, we establish this methodology by determining a spin gap of $\sim0.6\,\mu$eV in Na$_{2}$Co$_2$TeO$_{6}$.

\begin{figure}[!b]
\centering
\includegraphics[width=\columnwidth]{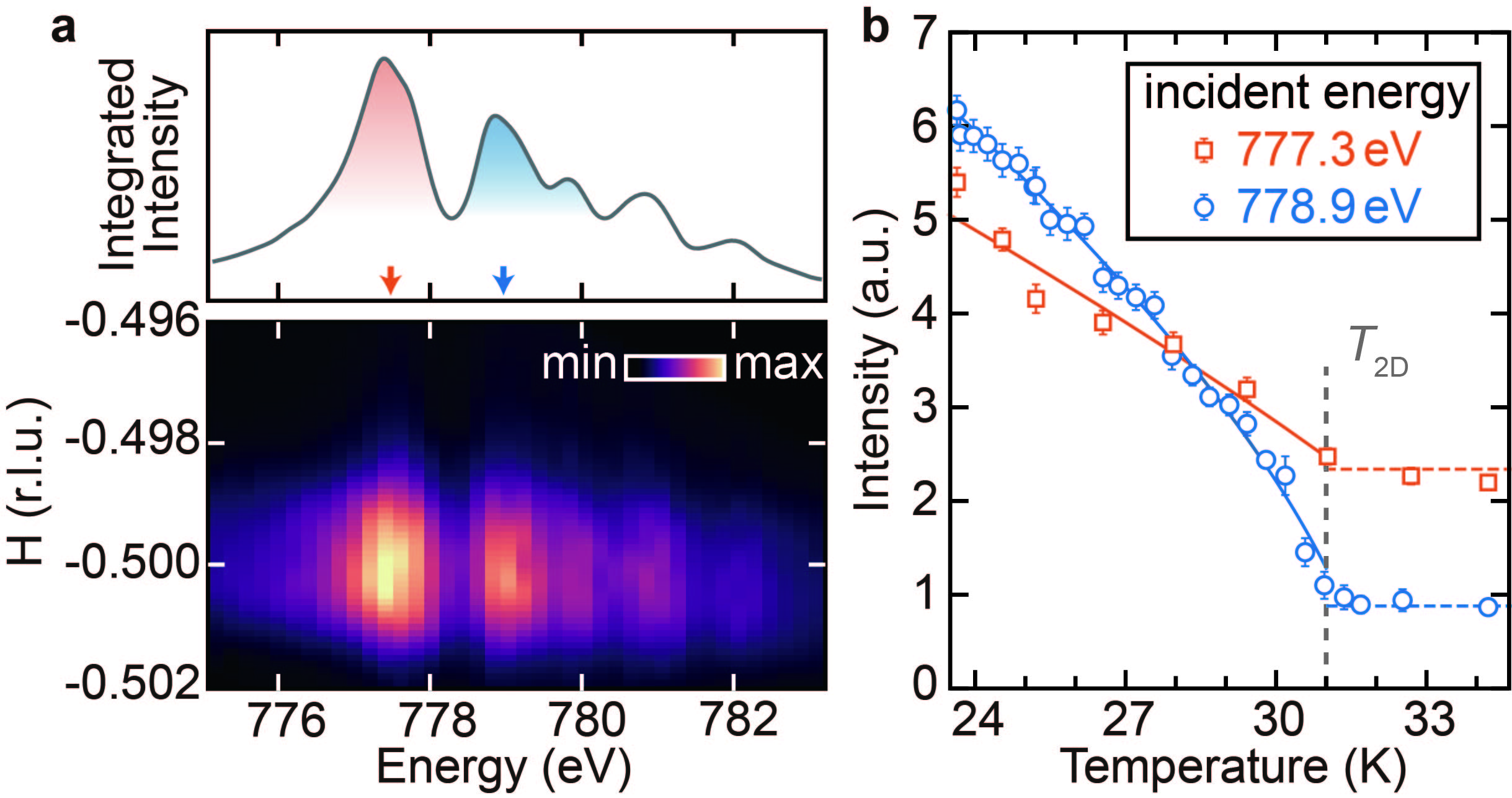}\vspace{-2mm}
\caption{{\bf Equilibrium characterization of magnetic order.} \textbf{a} REXS spectra for Na$_{2}$Co$_2$TeO$_{6}$ at 30\,K, showing the scattering intensity distribution around $H$\,=\,-0.5\,r.l.u. (lower) and the integrated intensity for wavevectors between $H$\,=\,-0.502 and -0.498\,r.l.u. (upper). The two shaded peaks in the upper panel highlight the two dominant resonant energies (777.3\,eV and 778.9\,eV) associated with the magnetic order.  \textbf{b} Temperature dependence of the scattering signals at the two different incident energies. The intensities are fitted using an empirical equation  
and revealed the 2D magnetic transition temperature $T_{\rm 2D}=31$\,K.
\label{fig:equilibrium}}
\end{figure}

\section{Results}

\subsection{Spectral Characterization in Equilibrium}

The Na$_{2}$Co$_2$TeO$_{6}$ crystal is characterized by a hexagonal non-centrosymmetric space group $P6_{3}22$. Its magnetic moments are predominantly contributed by the valence electron in the high-spin electronic configuration ($t^{5}_{2g}$$e^{2}_{g}$) of Co$^{2+}$ ions, with both spin and orbital angular momenta contributing to the magnetic moment\,\cite{PhysRevB.101.085120}. Each edge-sharing CoO$_6$ octahedra can be effectively regarded as a spin-$1/2$ state, forming a two-dimensional (2D) honeycomb lattice, as depicted in Fig.~\ref{fig:cartoon}\textbf{b}.  The energy-momentum-resolved REXS spectrum exhibits a pronounced peak at $\mathbf{q} = (-0.5, 0, 0.62)$ at 30\,K, as shown in Fig.~\ref{fig:equilibrium}$\textbf{a}$, consistent with previous study\,\cite{PhysRevB.103.L180404,PhysRevResearch.5.L022045}. Its intensity maximizes at two incident x-ray energies of 778.9\,eV and 777.3\,eV. As the temperature decreases below 2D magnetic phase transition temperature $T_{2D}\sim31$\,K, both peaks exhibit a coincided rise of intensity [see Supplementary Note 1 for details], reflecting their associations with the magnetic order\,\cite{PhysRevB.103.L180404, PhysRevB.103.214447}. Both scattering peaks remain finite while largely suppressed above 31\,K, indicating a subleading structural order that coexists with magnetic instability. This superstructural order likely originates from the Na atomic layers\,\cite{PhysRevB.103.L180404}. 
Analyzing the temperature dependency of these two scattering peaks facilitates the differentiation between magnetic and structural contributions. The scattering intensity at 778.9\,eV displays a strong temperature dependence immediately below 31\,K, indicating a predominant magnetic contribution; conversely, the intensity at 777.3\,eV remains evident above 31\,K and climbs more gradually below 31\,K, implying larger structural contribution. Our time-resolved experiment results show that the pump laser is not able to melt the superstructure diffraction [see Supplementary Note 2].

\begin{figure}[!t]
\centering
\includegraphics[width=\columnwidth]{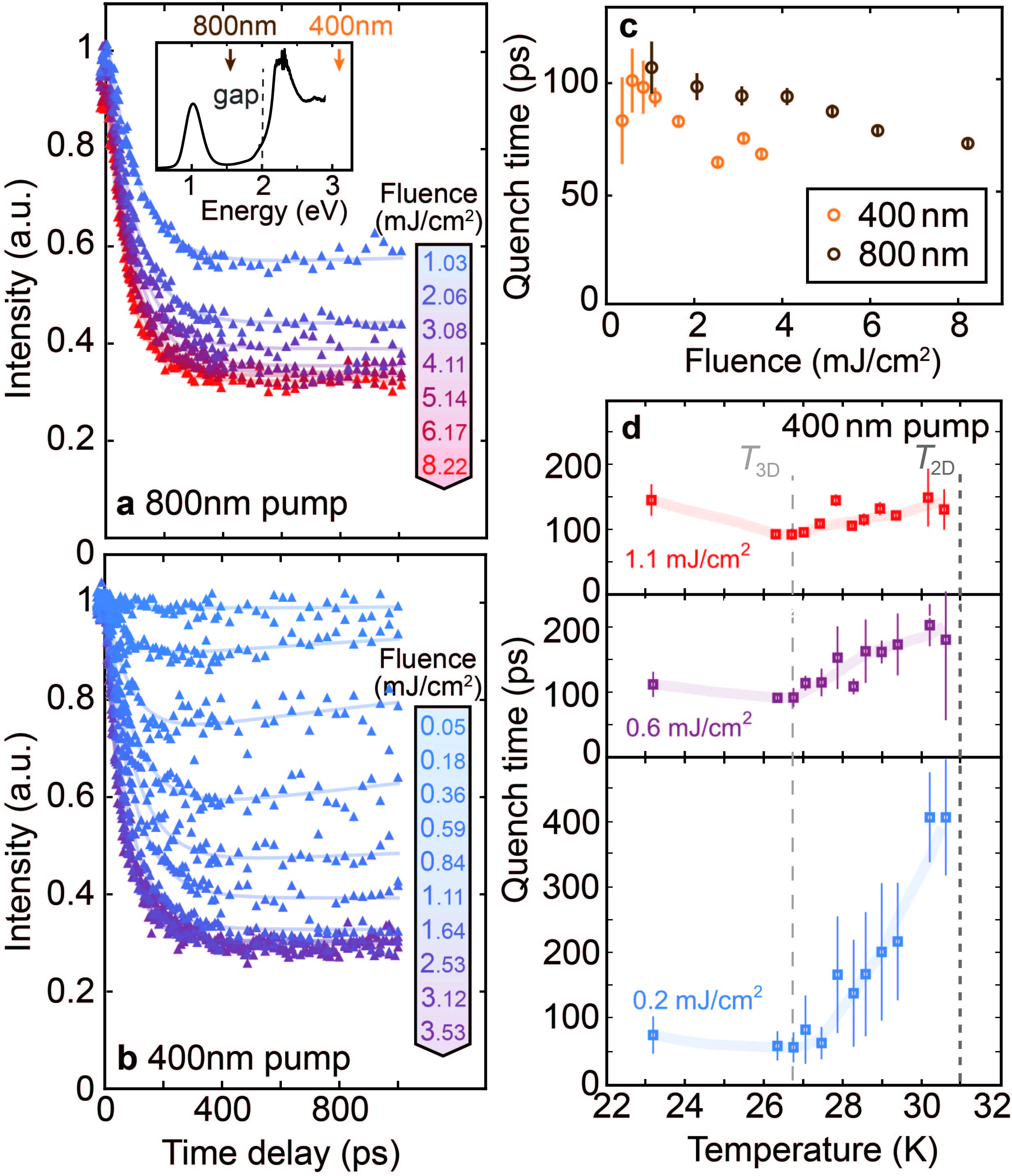}\vspace{-3mm}
\caption{{\bf Evolution of the magnetic scattering peak.} \textbf{a,b} Evolution of the magnetic scattering intensity for the 778.9\,eV resonance induced by \textbf{a} 800\,nm and \textbf{b} 400\,nm pumps with various fluences, normalized by the equilibrium spectral intensities. The solid curves delineate the double-exponential fitting using Eq.~\eqref{eq:doubleExpFitting}. The pump-probe measurements are conducted at 23\,K. The inset shows the optical absorption spectra of Na$_{2}$Co$_2$TeO$_{6}$, with arrows highlighting the 400\,nm and 800\,nm photon energies, respectively. \textbf{c} The quench time $\tau_{\rm q}$ extracted from \textbf{a} and \textbf{b} as a function of pump fluence. \textbf{d} Temperature dependence of the quench times for a 400\,nm pump with different pump fluences. The dashed lines denote the transition temperatures of the 3D and 2D magnetic orders. 
\label{fig:dynmFluence}}
\end{figure}

\subsection{Light-Induced Dynamics of the Magnetic Scattering Peak}

By driving Na$_{2}$Co$_2$TeO$_{6}$ out of equilibrium using a laser pulse, we analyze the subsequent changes in the magnetic structure using Tr-REXS. We exploit the two resonant scattering peaks (highlighted in Fig.~\ref{fig:equilibrium}) near $\mathbf{q} = (-0.5, 0, 0.62)$ to quantify the magnetic excitations. The energy gap of Na$_{2}$Co$_2$TeO$_{6}$ is determined as $\sim$2 eV by the optical absorption spectrum (inset of Fig.~\ref{fig:dynmFluence}$\textbf{a}$). The small absorption peak at $\sim$1\,eV arises from the d-d transition of octahedral Co$^{2+}$\,\cite{absorption}. 
To discern spin dynamics from those stemming from charge fluctuations, we examine both the in-gap pump with 800\,nm ($\sim$1.55\,eV) laser and cross-gap pump with 400\,nm ($\sim$3.1\,eV) laser. The cross-gap pump triggers charge excitations, while the in-gap pump largely does not, indicating that the similar dynamic features observed under both conditions predominantly stem from magnetic excitations.

As shown in Fig.~\ref{fig:dynmFluence}$\textbf{a}$, the evolution of the 778.9\,eV REXS peak intensity, induced by an 800\,nm pump laser, exhibits a prolonged dynamics at 23\,K, manifesting as a gradual decay after pump (denoted as ``quenching'') and a subsequent slow recovery back to equilibrium. This behavior can be described by a double-exponential function [see fitting details in Supplementary Note 3]:
\begin{equation}\label{eq:doubleExpFitting}
I(\Delta t) = I_{\rm res} + I_{\rm quench}\left[e^{-{{\Delta}t}/{\tau_{\rm q}}} +(1-e^{-{{\Delta}t}/{\tau_{\rm r}}})\right]\,.
\end{equation}
Here, $I_{\rm quench}$ is the intensity portion affected by the pump laser and depends on the pump fluence, while $I_{\rm res}$ represents the residual intensity. To highlight the relative changes, these intensities are normalized by the equilibrium scattering peak intensity. 
The two prolonged dynamical processes are characterized by the characteristic quenching time $\tau_{\rm q}$ and characteristic recovery time $\tau_{\rm r}$, respectively. Similar slow dynamics also appear in the evolution induced by a cross-gap pump with a 400\,nm laser (see Fig.~\ref{fig:dynmFluence}$\textbf{b}$). This similarity suggests that the dynamics induced by these long-term dynamics can be regarded as the evolution of spin excitations. Due to the optical resonance, the absorption rate is much higher in 400\,nm, resulting in a more pronounced response and higher fidelity when fitting the characteristic times [see Supplementary Note 4]. Specifically, the $I_{\rm quench}$ saturates at a pump fluence of 2.5\,mJ/cm$^2$ and 5\,mJ/c${\mathrm{m}}^{2}$ for the 400\,nm and 800\,nm pumps, respectively. For convenience of analyzing the magnetic excitation timescales and their fluence dependence, we primarily focus on the 400\,nm pump in this work.

The light-induced demagnetization in Na$_{2}$Co$_2$TeO$_{6}$ occurs within a timescale of 50 to 100\,ps, as characterized by $\tau_{\rm q}$ (see Fig.~\ref{fig:dynmFluence}$\textbf{c}$). This demagnetization process slightly accelerates with an increase in pump fluence, attributed to photocarrier screening effects\,\cite{a-RuCl3Wagner,Ta2NiSe5decayfluence}. Notably, the order parameter decays significantly slower than in CDW materials\,\cite{doi:10.1126/science.1160778,PhysRevLett.95.117005,PhysRevLett.123.097601,AlfredZong2018CDW,AlfredZong2019CDW} and Mott insulators\,\cite{PhysRevLett.106.217401, PhysRevB.96.184414, versteeg2020nonequilibrium,dean2016ultrafast,Sr3Ir2O7}. This phenomenon can be presumably attributed to the localized spins in cobalts, which hinder direct coupling between the electronic orbital degree of freedom and the spins. Without this direct coupling, the demagnetization is primarily driven by the strong phonon-magnon coupling in Na$_{2}$Co$_2$TeO$_{6}$ as also evidenced by thermal conductivity\,\cite{PhysRevB.104.144426,hong2023phonon}. 
This mechanism aligns with observations in other materials like InMnAs\,\cite{PhysRevLett.95.167401} and MnBi$_{2}$Te$_{4}$\,\cite{MnBi2Te4}, where localized spin moments and significant spin-lattice coupling are present. Similar to the REXS peak at 778.9\,eV, the dynamics of the 777.3\,eV resonant peak exhibit a slow demagnetization process with comparable timescales [see Supplementary Note 5]. Due to the mixture of magnetic and structural contributions, the $I_{\rm quench}$ is smaller for the 777.3\,eV REXS peak.

Aside from the similarity with the dynamics induced by the in-gap 800\,nm laser, relation between the prolonged dynamics induced by the 400\,nm laser and spin excitations is further confirmed through the temperature dependence. As shown in Fig.~\ref{fig:dynmFluence}$\textbf{d}$, the quenching time $\tau_{\rm q}$ exhibits a significant increase as the temperature approaches $T_{\rm 2D}=31$\,K from below. This trend reflects the strong magnetic fluctuations near the phase transition\,\cite{PhysRevX.9.021020,magneticfluctuation,versteeg2020nonequilibrium,PhysRevB.102.115143,collins1989magnetic}, which has been also observed in other Kitaev materials such as $\alpha$-RuCl$_3$\,\cite{versteeg2020nonequilibrium}. A minor difference from $\alpha$-RuCl$_3$ is the disruption of the monotonic trend below $26.7$\,K in Na$_{2}$Co$_2$TeO$_{6}$, where another 3D magnetic order starts to develop. This observation suggests an interplay between two distinct magnetic orders in Na$_{2}$Co$_2$TeO$_{6}$, a phenomenon that warrants further investigation but is beyond the scope of this work.

\begin{figure}[!t]
\centering\includegraphics[width = \columnwidth]{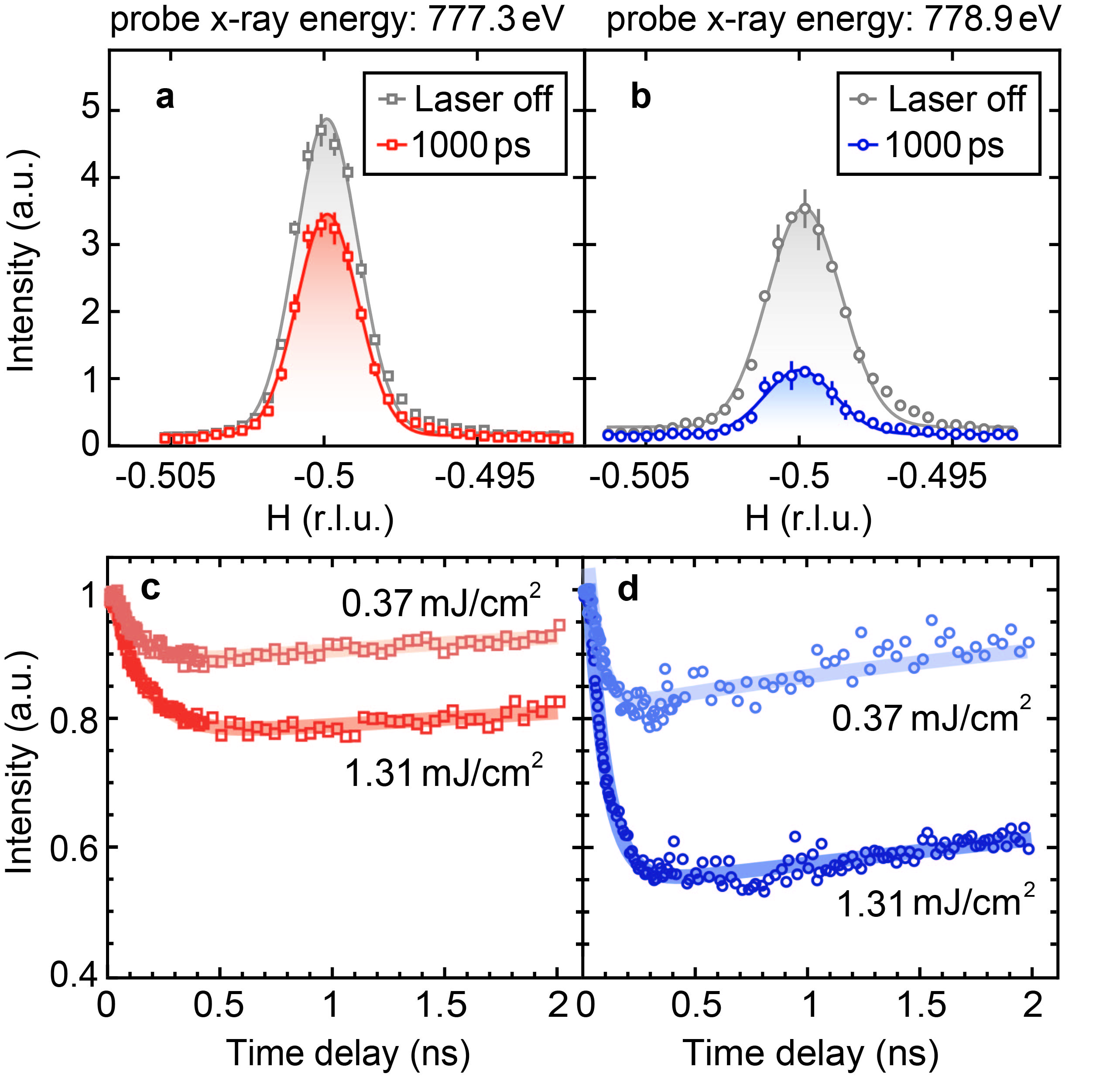}\vspace{-3mm}
\caption{\label{fig:phasediagram}\textbf{Destruction and recovery of  magnetic diffraction with a 400\,nm excitation.} 
\textbf{a,b} The variation of the (-0.5, 0, 0.62) magnetic peak before and after pump excitation (1000\,ps, 400\,nm) with 777.3\,eV and 778.9\,eV at 23\,K. The blue shade indicates the change in intensity caused by photoexcitation. Solid lines are fit using a Gaussian function. Error bars represent Poisson counting error. \textbf{c,d} Magnetic intensity normalized by laser-off data as a function of time delay with 777.3\,eV \textbf{c} and 778.9\,eV \textbf{d} at 23\,K. The blue and red lines display the results of fitting, corresponding to pump fluence of 0.37 and 1.31\,mJ/c${\mathrm{m}}^{2}$ respectively. 
\label{fig3}}
\end{figure}

Unlike the quench time which reflects the transition to high-energy excited states, the recovery process back to equilibrium characterizes the low-energy properties. Therefore, we examine the long-time evolution of the two Tr-REXS peaks at $\mathbf{q}= (-0.5, 0, 0.62)$ following the 400\,nm pump. A comparison between the spectral shapes for both resonant peaks at equilibrium and 1000\,ps indicates that the correlation length ($\sim$ 100\,$a_0$ and 600\,\AA) remains essentially consistent, distinct from the broadening observed in thermal fluctuations\,\cite{CLLSNO} (see Fig.~\ref{fig3}$\textbf{a, b}$ and Supplementary Note 6). 
Figures~\ref{fig3}$\textbf{c}$ and $\textbf{d}$ present the corresponding fluence dependence of the magnetic dynamics for the two x-ray energies. 
For a relatively strong pump of  1.31\,mJ/c${\mathrm{m}}^{2}$ , the long-term recovery can be characterized as $\tau_{\rm r}=8.66\pm0.47$\,ns for the 777.3\,eV and 11.43 $\pm$ 0.38\,ns for the 778.9\,eV peak. Reducing the pump fluence leads to a decrease in $\tau_{\rm r}$, contradicting again to the heat diffusion process [see Supplementary Note 7 for details]. At a fluence of 0.37\,mJ/c$\mathrm{m}^2$, we observe a saturation in the recovery timescale for the 778.9\,eV peak [see the Supplementary Note 7], yielding a $\tau_{\rm r}=2.54\pm0.14$\,ns. This timescale coincides with the results for the 777.3\,eV peak at the same fluence ($\tau_{\rm r}=3.94\pm0.17$\,ns). Their proximity further underlines their association with the material's intrinsic low-energy states, independent of pump and probe conditions. The phenomenon of prolonged relaxation is not unique to Na$_{2}$Co$_2$TeO$_{6}$, but is also observed in other materials with intricate magnetic phases, such as another Kitaev candidate $\alpha$-RuCl$_{3}$\,\cite{a-RuCl3Wagner} and multiferroic TbMnO$_{3}$. These nanosecond-long relaxations are believed to occur due to nearly degenerate magnetic ground and excited states, influenced by frustrated interactions\,\cite{lefranccois2016magnetic, bera2017zigzag, PhysRevB.103.L180404,PhysRevLett.129.147202,PhysRevB.103.214447}. Therefore, the timescale of recovery quantitatively encodes information about these low-energy states.

\subsection{Spin Gap Measurement from the Heisenberg-Kitaev model}

\begin{figure}[!t]
\centering
\hspace{-5mm}
\includegraphics[width=9cm]{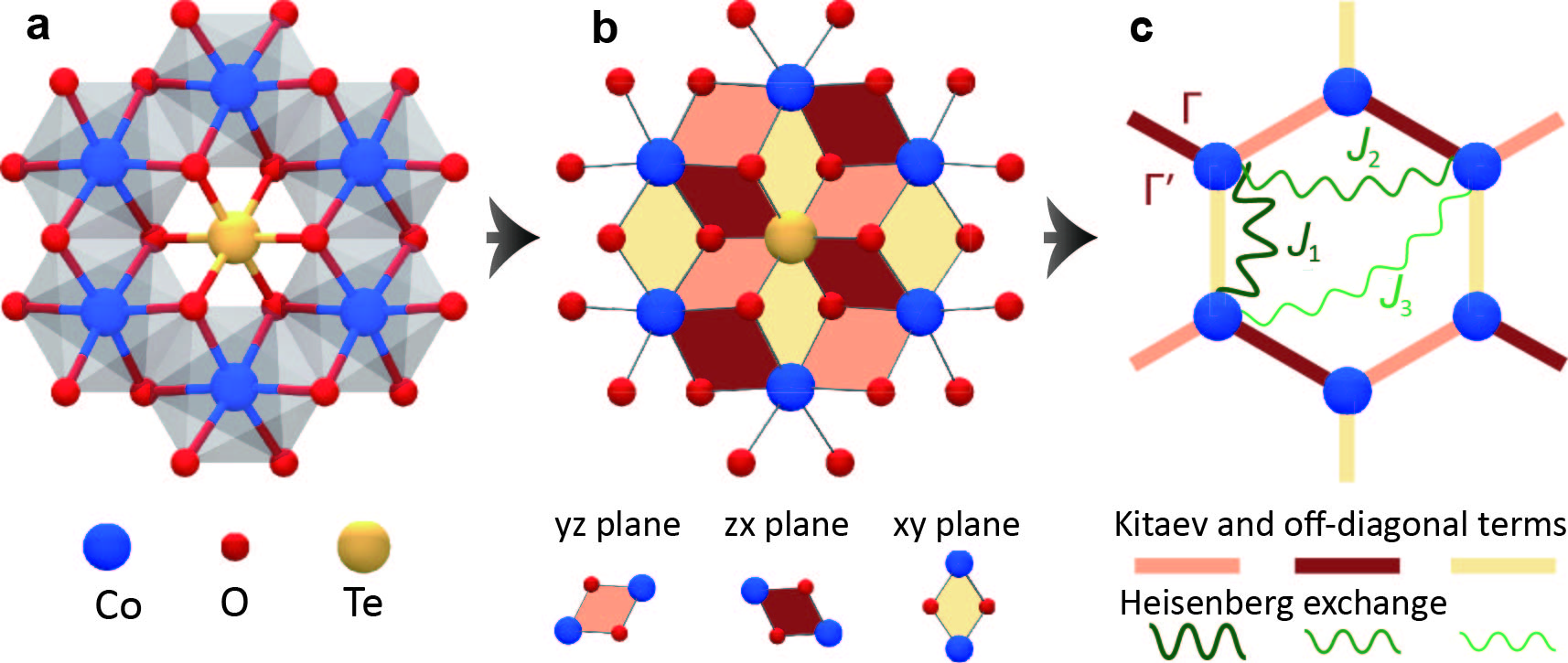}\vspace{-2mm}
\caption{\label{fig5}\textbf{Lattice structure of Na$_2$Co$_2$TeO$_6$ and the microscopic model schematics on the honeycomb layer.} \textbf{a} Co-Te-O layer with Co (blue) atoms forming the honeycomb structure. Each nearest-neighbor Co pair is connected through two oxygen (red), and the Te (yellow) is placed at the center of each hexagon. The edge-sharing octahedra are represented by the gray cage surrounding each Co. \textbf{b} The pseudospin superexchange of Co mediated by O gives rise to the anisotropic spin terms $K$, and the octahedra distortion results in the symmetric off-diagonal terms \{$\Gamma$, $\Gamma'$\}. The rotations of $\{\alpha, \beta, \gamma\}$ are represented by $\{y, z, x\}$ (orange), $\{z, x, y\}$ (red) and $\{x, y, z\}$ (yellow), respectively. \textbf{c} The isotropic Heisenberg interaction between the nearest neighbor, next nearest neighbor, and third nearest neighbor for Co are shown as \{$J_1$, $J_2$, $J_3$\}, the orientations of Kitaev and off-diagonal spin interactions are shown as the orange, red, and yellow bonds.\vspace{-2mm}}
\end{figure}

To elucidate the slow recovery dynamics in Na$_{2}$Co$_2$TeO$_{6}$ and its relation to the low-energy spin structure, we simulate the spin-$1/2$ Kitaev-Heisenberg Hamiltonian to describe the microscopic states of Na$_{2}$Co$_2$TeO$_{6}$\,\cite{songvilay2020kitaev,Winter_2022}. This model captures the superexchange between each Co atom in a hexagonal structure, as illustrated in Fig.~\ref{fig5}$\textbf{a}$. The cobalt atoms in an octahedral crystal field assume a high-spin $(t_{2g})^5(e_g)^2$ configuration in their $3d$ orbitals. Without trigonal splitting, an effective orbital momentum $L_{\text{eff}}=1$ with spin-orbit coupling splits the three-fold degenerate $t_{2g}$ orbitals into $j_{\text{eff}}=1/2, 3/2$, and $5/2$ states, with $j_{\text{eff}}=1/2$ as the ground state. The superexchange interactions between nearest-neighbor Co atoms, via $t_{2g}$--$t_{2g}$, $t_{2g}$--$e_g$, and $e_g$--$e_g$ channels, are mediated by the $2p$ orbitals of intervening oxygen atoms in a $90^{\circ}$ of Co--O--Co bond geometry [see Fig.~\ref{fig5}$\textbf{b}$]. This configuration leads to a coupling between the spins at two Co sites (denoted as $i$ and $j$) perpendicular to the exchange path, i.e.~$S^\gamma_i S^\gamma_j$ ($\gamma = {x, y, z}$ depends on the hexagon plane orientation). 
Deviations from perfect octahedral geometry introduce off-diagonal spin interactions, resulting in the Kitaev-Heisenberg Hamiltonian:
\begin{eqnarray}
\mathcal{H}&=& \sum_{\langle i,j\rangle} J_1 ~\mathbf{S}_i\cdot\mathbf{S}_j + \sum_{\langle\!\langle i,j\rangle\!\rangle} J_2 ~\mathbf{S}_i\cdot\mathbf{S}_j + \sum_{\langle\!\langle\!\langle i,j\rangle\!\rangle\!\rangle} J_3 ~\mathbf{S}_i\cdot\mathbf{S}_j\nonumber\\
&&+ \sum_{\langle i,j\rangle} K S^\gamma_i S^\gamma_j +  \sum_{\langle i,j\rangle} \Gamma \left(S^\alpha_i S^\beta_j + S^\beta_i S^\alpha_j\right) \nonumber\\
&&+  \sum_{\langle i,j\rangle} \Gamma^\prime \left(S^\alpha_i S^\gamma_j + S^\gamma_i S^\alpha_j+S^\beta_i S^\gamma_j + S^\gamma_i S^\beta_j\right)\,.\label{hamiltonian}
\end{eqnarray}
Here, the Heisenberg interactions are described by the first, second, and third nearest-neighbor spin-exchange $J_1$, $J_2$, and $J_3$; $K$ is the bond-dependent Kitaev interactions, with $\{\alpha,\beta,\gamma\}$ denoting the three types of anisotropic terms for the three nearest-neighbor directions. The symmetric off-diagonal terms, $\Gamma$ and $\Gamma^\prime$, appear in the Hamiltonian due to the octahedra distortion mentioned above. We follow the spectral fitting in Ref.~\onlinecite{songvilay2020kitaev} and choose the coupling coefficients as $J_1=-0.1$\,meV, $J_2=0.3$\,meV, $J_3=0.9$\,meV, $K=-9$\,meV, $\Gamma=1.8$\,meV, and $\Gamma^\prime=0.3$\,meV. 

\begin{figure}[!t]
\centering
\includegraphics[width=9cm]{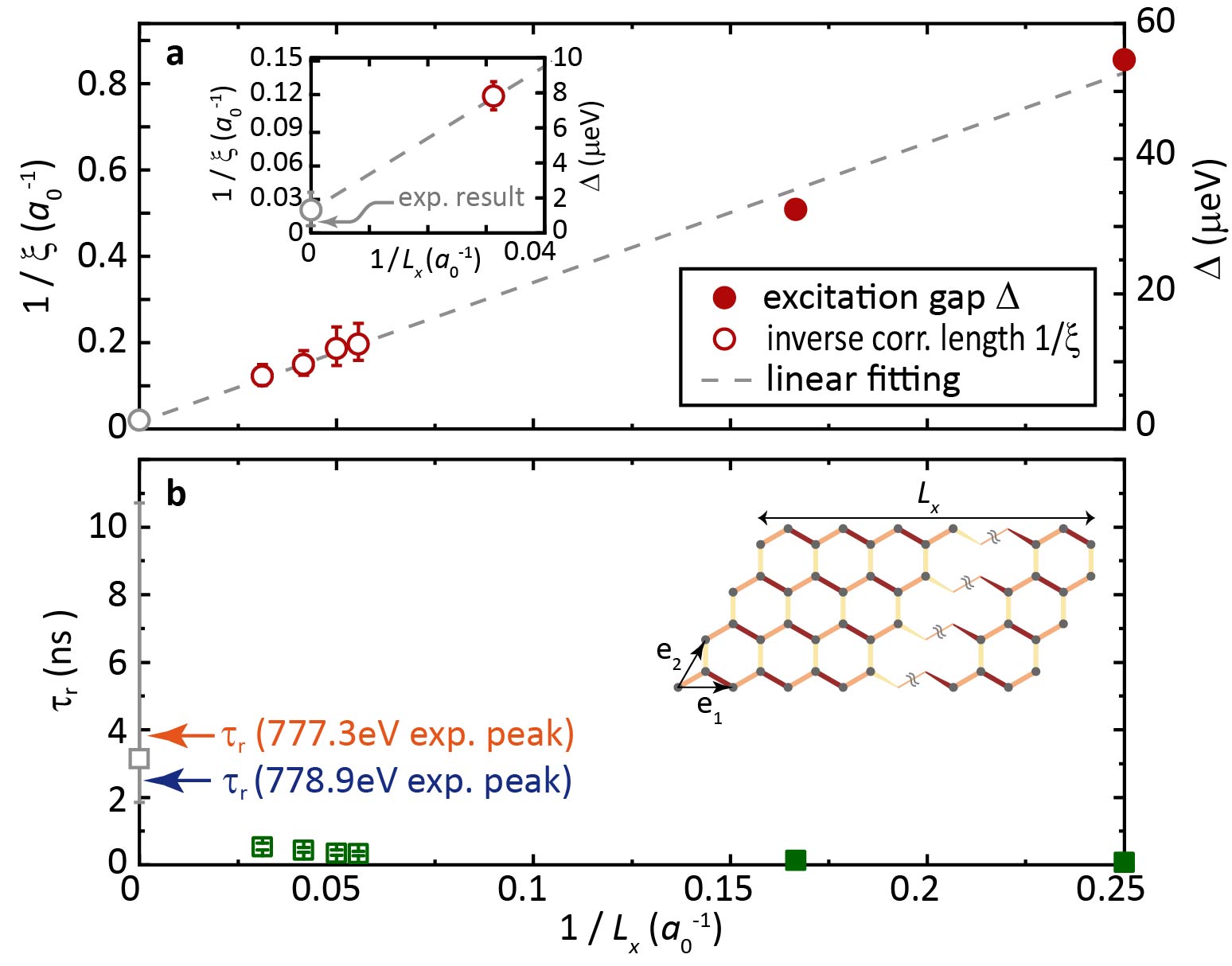}\vspace{-3mm}
\caption{\label{fig6}\textbf{Simulation of energy and timescales using the Kitaev-Heisenberg model.} \textbf{a} The excitation gap $\Delta$ (closed circles) and inverse correlation length $1/\xi$ (open circles) simulated using the Kitaev-Heisenberg model for $L_x\times 4$-systems with different sizes. The dashed line denotes the linear fitting for the $\Delta$ and $1/\xi$, with the thermodynamic-limit extrapolation (intersection) represented by the gray open circle. The inset contrasts the simulated and experimental correlation lengths near the thermodynamic limit. \textbf{b} Characteristic timescales $\tau_{\rm r}$ (green squares) extracted from the Rabi oscillation across the excitation gap. The gray square marks the thermodynamic-limit extrapolation. The two arrows highlight experimentally determined $\tau_{\rm r}$ using two x-ray energies. The inset shows the quasi-1D geometry, with bond colors indicating the Kitaev superexchanges in Fig.~\ref{fig5}. }
\end{figure}

Employing the DMRG method\,\cite{White1992}, renowned for its accuracy in analyzing strongly correlated systems, we simulate the electronic structure of Na$_{2}$Co$_2$TeO$_{6}$, with a specific focus on the spin excitation gap. For relatively small systems, DMRG enables direct calculation of the first several excited states, thereby identifying the spin excitation gap defined as the energy difference between the ground state and the first excited state. The deduced gap values for a $4\times4$ and $6\times4$ clusters are $\Delta =54.6$\,$\mu$eV and $32.4$\,$\mu$eV, respectively. (These unit cells are oriented along the $\mathbf{e}_1$ and $\mathbf{e}_2$ vectors, as shown in the inset of Fig.~\ref{fig6}$\textbf{b}$.) Notably, the gap decreases as the system size increases, indicating a finite-size effect and necessitating extrapolation to the thermodynamic limit. 

Despite the impracticality of a direct finite-size scaling for excited states due to computational demands, we employ an indirect extrapolation to determine the spin gap. Systems larger than $18\times4$ exhibit a spatial distribution of the ground-state spin-spin correlation function $F^z(r)$ that decay exponentially along the horizontal direction. A correlation length $\xi$ can be extracted from fitting the correlation function (see the \textbf{Methods}). Accounting for relativistic effects near a quantum phase transition, we employ the $\Delta \propto 1/\xi$ relationship to infer the spin gap\,\cite{eberharter2023extracting}. Specifically, we simulate the $F^z(r)$ for systems with various $L_x$ where $\xi$ can be reliably determined. The inverse of these $\xi$-values displays strong linearity with $1/L_x$, as shown in Fig.~\ref{fig6}$\textbf{a}$. Extrapolating this size dependency to the thermodynamic limit ($1/L_x\rightarrow0$), we obtain an intersection of $1/\xi\sim0.02a_0^{-1}$, with $a_0$ representing the lattice constant. This extrapolated result is consistent with experimentally obtained $1/\xi\sim0.01a_0^{-1}$, without any parameter tuning of the simple model. By comparing the fitted linearity with the energy gaps calculated for smaller systems, we establish a correlation between $\Delta$ and $1/\xi$ (velocity of excitations), yielding $\xi\Delta/a_0 = 63.9\pm4.5\,\mu$eV. Therefore, our DMRG simulation of the Kitaev-Heisenberg gives a spin gap $\Delta\sim1.3~ \mu$eV in the thermodynamic limit. 

The long-term recovery timescale is predominantly governed by the low-energy states and can be approximated by the Rabi oscillation period across the spin gap. Therefore, the characteristic time derived from our simulation yields $\tau_{\rm r} = h/\Delta\sim3.2$\,ns, as shown in Fig.~\ref{fig6}$\textbf{b}$. This simulated timescale is consistent with the experimentally observed $\tau_{\rm r}$ at low frequencies for both x-ray peaks (3.94\,ns and 2.54\,ns).

\section{Discussions}

The consistency between the spin gap determined through DMRG simulations and the characteristic recovery time identified in trREXS experiments reflects the viability of measuring ultra-small energy gaps using pump-probe x-ray experiments. Importantly, such determinations are feasible without the intervention of theoretical simulations. A supporting evidence is that the experimentally measured correlation length ($\xi=100a_0$) translates into an energy gap $\Delta\sim 0.6\,\mu$eV, using the $\xi\Delta/a_0$ obtained in Fig.~\ref{fig6}$\textbf{a}$, yielding a characteristic time of $6.4$\,ns. This timescale falls within the range of the recovery times ($\tau_{\rm r}$) observed for the strong and weak pump conditions. While our study utilizes Na$_{2}$Co$_2$TeO$_{6}$ as an example, due to the significance of its spin gap, this methodology is broadly applicable to diverse systems where the gap size ranges from 0.1 to 100\,$\mu$eV. Such gap sizes, elusive to the resolution of INS, correspond to picosecond to nanosecond timescales, which can be effectively revealed by the recovery dynamics.

While the correlation length and associated timescale exhibit consistent orders of magnitude, it is important to emphasize that the Hubbard-Kitaev model, as a single-orbital spin model, simplifies the material by focusing on the key ingredients. It does not account for charge transfer between different atoms, the presence of 3D magnetic order, or spin-phonon couplings.  As a result, the model does not capture the demagnetization process observed in the initial 100 ps of the dynamics. The DMRG simulation is designed to examine the recovery process, governed by the low-energy states. Moreover, experimental results are subject to fitting inaccuracies and the intrinsic noise prevalent in long-duration measurements. As such, when comparing simulation outcomes with experimental data, one should consider potential error cancellations and focus on the order of magnitude, avoiding over-interpretation.

Our findings contribute to the ongoing exploration of spin gaps in QSLs and other complex magnetic systems. Small spin gaps have been frequently reported in QSL candidates through various measurements. For example, thermal conductivity assessments have suggested a spin gap of $\sim$30\,$\mu$eV in triangular lattice $\kappa$-(BEDT-TTF)$_{2}$Cu$_{2}$(CN)$_{3}$\,\cite{thermalconductivity}. NMR measurements have characterized a $\sim$0.8\,meV spin gap of herbertsmithite ZnCu$_3$(OH)$_6$Cl$_2$\,\cite{gappedZn}. The significance of a spin gap lies in its sensitivity to non-Kitaev interactions \,\cite{PhysRevLett.117.037209,PhysRevLett.119.227202}, which hinder the realization of long-sought QSL. While materials like $\alpha$-RuCl$_{3}$ and Na$_{2}$Co$_2$TeO$_{6}$ may not perfectly exhibit QSL characteristics, accurately identifying their spin gaps helps potential design towards a Kitaev QSL region under specific conditions, such as an applied magnetic field\,\cite{ESRNCTO,PhysRevLett.120.117204,PhysRevLett.119.037201}. The demonstration of time-resolved x-ray scattering spectroscopy in detecting small spin gaps with unprecedented precision opens innovative pathways for the detailed study and engineering of quantum magnets.\\

\noindent
\section*{METHODS}
\subsection*{Sample preparation}

The high-quality single crystals of Na$_{2}$Co$_2$TeO$_{6}$ used in this study were grown with a flux method\,\cite{PhysRevB.101.085120}. The hexagonal-shaped crystal flake used in this study had the dimensions of $\sim$\,3\,mm$\times$3\,mm$\times$0.2\,mm (lattice parameters: $a$ = $b$ = 5.25\,\AA, $c$ = 11.19\,\AA). The sample was cleaved and examined by x-ray diffraction measurements (XRD) before experiments to confirm its excellent quality.
Single crystal x-ray diffraction measurements were performed using the custom-designed x-ray instrument equipped with a Xenocs Genix3D Mo K$\alpha$ (17.48\,keV) x-ray source, which provides $\sim$\,2.5\,$\times$\,$10^7$ photons/sec in a beam spot size of 150\,$\mu$m at sample position\,\cite{PhysRevResearch.5.L012032}.

\subsection*{Optical measurements}

The optical transmission data were collected at room temperature and converted to the absorption spectra. For the photon energy range from 0.5 to 2.7\,eV, the measurement was performed on a Bruker 80V Fourier transform infrared spectrometer. For the photon energy ranges from 1.3 to 4\,eV, the measurement was carried out on a home-built transmission measurement setup with a deuterium-halogen light source (Ideaoptics iDH2000-BSC) and a highly sensitive spectrometer (Ideaoptics Nova). Then, the transmission spectra from the two measurements were combined by normalizing the data in the overlapped range.

\subsection*{tr-RSXS measurements}

The tr-RSXS experiments were carried out at the SSS-RSXS endstation of PAL-XFEL\,\cite{jang2020time}. The sample was mounted on a six-axis open-circle cryostat manipulator with a base temperature of $\sim$20\,K. The sample surface was perpendicular to the crystalline c axis, and the horizontal scattering plane was parallel to the bc plane. X-ray pulses with $\sim$80\,fs pulse duration and 60\,Hz repetition rate were used for the soft x-ray probe. The x-ray was linear horizontal polarized ($\pi$-polarization), and the photon energy was tuned to Co L$_{3}$ edge ($\sim$778\,eV). Since the 2D magnetic order is $L$-independent, we fixed the scattering angle of detector at 2$\theta$ = 156$^{\circ}$, which provides $L$ = 0.62\,r.l.u. at $H$ = -0.5\,r.l.u.. 
The temperature dependence of the scattering signals was fitted by an empirical function $I(T) = a\left\{1-\left[(T+b)/(1+b)\right]^c\right\}$ as shown in Fig.~\ref{fig:equilibrium}$\textbf{b}$ \,\cite{EQfunction}. The equilibrium sample temperature had been calibrated following the magnetic transition temperature reported in Ref. \,\cite{PhysRevB.103.L180404}.

We utilized a Ti:sapphire laser to provide optical lasers at 1.55\,eV (800\,nm) and 3.1\,eV (400\,nm) with a pulse duration of $\sim$50\,fs and a repetition rate of 30\,Hz. Both $\sigma$-polarized (perpendicular to the scattering plane) and $\pi$-polarized (parallel to the scattering plane) laser pulses were used to excite Na$_{2}$Co$_2$TeO$_{6}$, showing similar transient responses, indicating no laser polarization dependence. To simplify our experimental setup, we chose an 800\,nm laser with $\sigma$-polarization and a 400\,nm laser with $\pi$-polarization, considering that the polarization direction of the laser pulse underwent a 90-degree rotation after frequency doubling. The overall time resolution was $\sim$\,108\,fs, determined by measuring the pump-probe cross-correlation. The optical laser was nearly parallel to the incident X-ray beam, with an angle difference of less than $1^\circ$. The pump fluence ranged from 0.1 to 8\,mJ/c${\mathrm{m}}^{2}$. The x-ray spot size at the sample position was $\sim$100 (H) $\times$ 200 (V)\,$\mu{\mathrm{m}}^{2}$ (FWHM), while the optical laser spot diameter was about $\sim$500\,$\mu$m (FWHM). The X-ray repetition rate was twice that of the pump pulses, enabling the comparison of diffraction signals before and after pump excitation. \\

\subsection*{Extrapolation of the Excitation Gap}
To estimate the gap in the thermodynamic limit, we simulated the first excited-state energy for the Heisenberg-Kitaev model on a four-leg ladder using excited-state method of DMRG developed on ITensor Software Library\,\cite{10.21468/SciPostPhysCodeb.4}. However, due to the significantly increased computing costs after extending the cylinder's length $L_x$, our excited-state simulations were restricted to $L_x=4$ and $6$ (in units of $a_0$). In simulations of larger systems, we switched to an indirect simulation of the gap using ground-state properties obtained from DMRG~\cite{Peng2021}. Specifically, we estimated the ground-state spin-spin correlation length using a code based on a high-performance matrix product state algorithm library GraceQ/MPS2\,\cite{GraceQ}. The number of DMRG block states was constrained owing to the absence of spin-rotational symmetry. We maintained multiple bond dimensions of the matrix product state representation of the ground state at each sweep, with a maximum bond dimension of 2048, and a typical truncation error of $\epsilon \approx 10^{-7}$. The results presented in the main text had been extrapolated to $\epsilon = 0$ to minimize the cutoff error. Note that the cylinder geometry, chosen as $L_x\times L_y$ cluster with the open boundary condition along the $\mathbf{e}_1$ direction and periodic boundary condition along the $\mathbf{e}_2$ direction, destroyed the continuous spin symmetry in the system so that the spin correlation function $F^\alpha(r)=\langle S^\alpha_{x_0}S^\alpha_{x_0+r}\rangle-\langle S^\alpha_{x_0}\rangle \langle S^\alpha_{x_0+r}\rangle$ for $\alpha=x$ and $y$ decayed exponentially, but $F^y(r)$ saturated to a constant while the spin gap was finite. We obtained the spin correlation length by fitting the correlation function $F^z(r)$ through $e^{-r/\xi}$. The extrapolation into the thermodynamic limit yields a $\xi=49a_0$. 

The correlation length was extrapolated to the thermodynamic limit $L_x\rightarrow\infty$, using the linear function $1/\xi = \alpha(1/L_x) + \beta$. This fitting function faithfully described the scaling behavior, reaching the fitted intercept $\beta=0.02059 ± 0.01348$. Here, the $\xi$ was expressed in the unit of unit cell along the $\mathbf{e}_1$ direction, depicted as $a_0$. As the linear relation between $1/\xi$ and $1/L_x$ had been verified, we reversely extrapolated the linear function backward to the short $L_x$ region where the energy difference between the ground state and the first excited state was evaluated through the excited-state method of DMRG. Then we used the least-square approach to determine the $\xi \Delta /a_0$ constant by minimizing the energy error against the two simulated excitation gaps for small clusters. The fitting result showed $\xi \Delta /a_0=63.9\pm 4.5\,\mu$eV, where the small residual error indicated the validity of this linear fitting. Alternatively, if we picked only one out of the $4\times 4$- or $6\times 4$-cluster to fit the constant, the estimation of $\xi \Delta /a_0$ would be $66.4\,\mu$eV or $58.4\,\mu$eV, respectively. This estimated error was consistent with the fitting residual calculated above. \\

\bibliographystyle{naturemag}
\bibliography{reference}

\vspace{1 ex}
\noindent
{\bf Acknowledgements:} We acknowledge the valuable discussion with Hong-Chen Jiang, Shaozhi Li, Donna N. Sheng, Yahui Zhang, and Alfred Zong. Cheng Peng acknowledges Gregory M. Stewart for the assistance in drawing FIG. 5. Y.Y.P. is grateful for financial support from the Ministry of Science and Technology of China (Grants No. 2021YFA140190 and No. 2019YFA0308401) and the National Natural Science Foundation of China (Grants No. 12374143 and No. 11974029). Y.W. acknowledges support by the U.S. Department of Energy, Office of Science, Basic Energy Sciences, under Early Career Award No.~DE-SC0024524. H.J. acknowledges the support by the National Research Foundation grant funded by the Korea government (MSIT) (grant no. 2019R1F1A1060295). The works at Max Planck POSTECH/Korea Research Initiative were supported by the National Research Foundation of Korea funded by the Ministry of Science and ICT, Grant No. 2022M3H4A1A04074153 and 2020M3H4A2084417. C.P. and J.J.T. acknowledge the support of the U.S. Department of Energy, Office of Science, Basic Energy Sciences under Award No. DE-SC0022216. This tr-RSXS experiment was performed at the SSS-RSXS endstation (proposal number: 2021-2nd-SSS-010) of the PAL-XFEL funded by the Korea government (MSIT). The computation for this research used resources of the National Energy Research Scientific Computing Center, a DOE Office of Science User Facility supported by the Office of Science of the U.S. Department of Energy under Contract No. DE-AC02-05CH11231. 
{\bf Author contributions:} 
Y.Y.P conceived and designed the experiments with suggestions from Y.L., H.J.; X.Y.J., Q.Z.Q., X.Q.C., Q.Z.L., H.J., S.Y.P., M.K., H.D.K and Y.Y.P. performed the tr-RSXS experiment at the PAL-XFEL with the help of B.L.. W.J.C., X.H.J. and X.Y.J. synthesized, grew and characterized the Na$_{2}$Co$_2$TeO$_{6}$ single crystals. L.Y., T.D. and N.L.W. carried out the optical transmission experiment and analyzed the data. Y.Y.P., X.Y.J. and Q.Z.Q. analyzed the tr-RSXS experimental data; C.P. conducted DMRG calculations with guidance and support from Y.W.; Y.Y.P., X.Y.J., J.J.T., C.P. and Y.W. wrote the manuscript with input and discussion from all co-authors. 
{\bf Competing interests:} The authors declare that they have no competing interests. {\bf Data and materials availability:} All data needed to evaluate the conclusions in the paper are present in the paper and/or the Supplementary Materials. Additional data related to this paper may be requested from the authors.

\end{document}